\documentstyle[12pt,epsf,rotate]{article}
\def\MeV{\,{\rm MeV}}

\def\Mpc{\,{\rm Mpc}}

\def\cmm2{{\,\rm cm^{-2}}}
\def\cm2{{\,{\rm cm}^2}}
\def\cmm3{{\,{\rm cm}^{-3}}}
\def\gcmm3{{\,{\rm g\,cm^{-3}}}}
\def\kms{\,{\rm km\,s^{-1}}}

\def\la{\mathrel{\mathpalette\fun <}}
\def\ga{\mathrel{\mathpalette\fun >}}
\def\fun#1#2{\lower3.6pt\vbox{\baselineskip0pt\lineskip.9pt
  \ialign{$\mathsurround=0pt#1\hfil##\hfil$\crcr#2\crcr\sim\crcr}}}

\begin{document}
\baselineskip=24pt
\pagestyle{empty}
\begin{center}
\bigskip

%\centerline{\Huge\bf Draft date}
\bigskip
\rightline{FERMILAB--Pub--95/407-A}
\rightline{astro-ph/9602050}
\rightline{submitted to {\it Nature}}

\vspace{.2in} {\Large \bf COSMOLOGICAL IMPLICATIONS OF THE FIRST
MEASUREMENT OF THE LOCAL ISM ABUNDANCE OF $^3$HE}
\bigskip

\vspace{.2in} Michael S. Turner,$^{1,2,3}$ James W. Truran,$^2$ David N.
Schramm,$^{1,2,3}$ and Craig J.~Copi$^{1,3}$\\

\vspace{.2in} {\it $^1$NASA/Fermilab Astrophysics Center\\ Fermi National
Accelerator Laboratory, Batavia, IL~~60510-0500}\\

\vspace{0.1in} {\it $^2$Department of Astronomy \& Astrophysics\\ Enrico
Fermi Institute, The University of Chicago, Chicago, IL~~60637-1433}\\

\vspace{0.1in} {\it $^3$Department of Physics\\ Enrico Fermi Institute, The
University of Chicago, Chicago, IL~~60637-1433}\\

\end{center}

\vspace{.3in}
\centerline{\bf ABSTRACT}
\bigskip

Deuterium plays a crucial role in testing big-bang nucleosynthesis. Its
chemical evolution, while simple (it is burned to $^3$He), is intertwined
with the more complicated evolution of $^3$He.
Gloeckler \& Geiss' new measurement of the $^3$He abundance and
the HST measurement of D, both in the local ISM today, can be
compared to the pre-solar nebula abundances of D and $^3$He.  Within the
uncertainties, the sum of D +
$^3$He relative to hydrogen is unchanged.  This provides some
validation of the cosmological utility of D + $^3$He, first suggested by
Yang et al (1984), and further, indicates that over the
past 4.5 Gyr there has been at most modest stellar production of $^3$He,
in contradiction with stellar modeling, or modest stellar
destruction of $^3$He, in contradiction with some ``solar spoons.''
While the earlier Galactic evolution of D + $^3$He cannot
be constrained directly, it is expected to be
dominated by massive stars, which deplete their $^3$He
and produce metals.  Based on the Galactic metallicity
and the constancy of D + $^3$He over the past 4.5 Gyr, we derive a
more empirically based lower bound to the cosmological baryon density; while
not dramatically different from the original bound of Yang et al (1984) based
on D + $^3$He, it alleviates some of the cosmic tension between
the big-bang $^4$He abundance and those of D and $^3$He.

\newpage
\pagestyle{plain}
\setcounter{page}{1}
\newpage

\section{Introduction}

Much of the current controversy concerning the consistency (Copi et al~1995a,
1995b) or inconsistency (Hata et al~1995a) of standard big-bang
nucleosynthesis revolves around the chemical evolution of $^3$He. 
In fact, $^3$He is involved indirectly.  Deuterium plays the
crucial role in testing big-bang nucleosynthesis,
as its abundance is the most sensitive to the baryon density, decreasing
rapidly with increasing baryon density, and its chemical evolution
brings in $^3$He.   The chemical evolution of D is
straightforward:  it is readily burned to $^3$He but it is not produced in
Galactic environments (Epstein, Lattimer \& Schramm 1976).  This means
that the present deuterium abundance can be  used to place an upper limit
to the baryon density.  This upper limit,
$\eta \la 9\times 10^{-10}$ which implies $\Omega_B \la 0.03h^{-2}\la 0.2$,
provides the linchpin in the two-decade-old argument that baryons cannot close
the Universe (Reeves et al~1973).  This argument is not questioned
in the current controversy.  (As usual, $\eta$ is the present ratio
of baryons to photons, $\Omega_B$ is the fraction of critical
density contributed by baryons, and the Hubble constant $H_0 =
100h\kms\Mpc^{-1}$ with $0.4<h<1$.)

Using deuterium to precisely determine the baryon density, or even to set
a lower limit to it, is more difficult.  Because deuterium is so easily
destroyed in passing through stars, the former is only possible if the
deuterium abundance can be measured in a very primitive sample of the
Universe.  While there has been much progress toward this goal, with
several detections and upper limits based on
the D Ly-$\alpha$ feature in high redshift ($z\sim
3$), absorption-line systems (York et al~1984; Songaila et al~1994;
Carswell et al~1985, 1995; Tytler \& Fann~1994; Rugers and
Hogan 1995; Wampler et al~1996) yielding inferred abundances in
the range from $2\times 10^{-5}$ to $2\times 10^{-4}$,
there is yet no definitive result.

The derivation of a lower limit to the baryon density hinges upon the
chemical evolution of $^3$He.  Since D is burned to
$^3$He and $^3$He is far more difficult to burn, Yang et al~(1984)
proposed using the sum of D + $^3$He for this purpose. Based upon stellar
modeling (Iben and Truran 1978), they assumed that at least 25\% of
the primordial D + $^3$He
survives stellar processing, which led to the lower limit
$\eta \ga 2.5\times 10^{-10}$ and $\Omega_B \ga 0.009h^{-2}$. This,
together with the upper limit that follows from $^7$Li ($\eta \la 6\times
10^{-10}$ and $\Omega_B\la 0.02h^{-2}$), provides the best determination of
the baryon density -- between about 1\% and 15\% of critical density
(allowing $0.4< h < 1$; see Copi et al~1995a) --  and establishes the two
dark-matter problems central to cosmology:  most of the baryons are dark
(since $\Omega_{\rm LUM} \simeq 0.003h^{-1} < \Omega_B$)
and most of the dark matter must be nonbaryonic, if
as several measurements indicate $\Omega_0 \ga 0.2$.

Beyond pinning down the baryon density, there is a more fundamental issue:
the consistency of the standard model of primordial nucleosynthesis itself
and the validity of the hot big-bang model at times as early as 0.01 sec.
(By standard model of big-bang nucleosynthesis we mean:  FRW cosmology,
uniform distribution of baryons, three light neutrino species, and small
neutrino chemical potentials.) The $^7$Li abundance measured in almost 100
old, Pop II halo stars is consistent (``at $2\sigma$'')
with the big-bang prediction provided
that $\eta \simeq (1-6)\times 10^{-10}$, which overlaps the D + $^3$He
consistency interval (Copi et al~1995a). Of some concern is the primeval
$^4$He abundance:  If one accepts at face value the analysis of Olive and
Steigman (1995), based upon metal poor, extragalactic HII regions, then
their value for $^4$He of $Y_P = 0.232 \pm 0.003\ {\rm (stat)} \pm 0.005\
{\rm (sys)}$ implies $\eta \simeq (1-4)\times 10^{-10}$ (at ``$2\sigma$'),
which is only marginally consistent with the D + $^3$He lower bound.
It should be
noted, however, that other authors (see e.g., Skillman et al~1994; Sasselov
and Goldwirth 1995; and Pagel, private communication)
have argued that the systematic uncertainties are at least
a factor of two larger, which, owing to the logarithmic dependence of $Y_P$
upon $\eta$ would enlarge the concordance interval to
$\eta \simeq (0.6 - 10)\times 10^{-10}$.

For some time, there has been tension between the measured abundances of $^4$He and
D + $^3$He (see e.g., Yang et al~1984; Copi et al~1995a; Walker et al~1991;
Olive et al~1995; Scully et al~1996).
The resolution could involve a revision of our
understanding of the evolution of $^3$He:
more stellar destruction than standard stellar models predict
would lead to a lower value of $\eta$ as inferred from D + $^3$He
and lessen the tension.  Alternatively, the resolution could
involve an underestimation of the primeval $^4$He abundance, by
an amount $\Delta Y_P
\sim 0.01$ (Copi et al~1995b); this would raise the
value of $\eta$ inferred from $^4$He, making it consistent
with that inferred from D + $^3$He and conventional stellar
evolution of $^3$He.  Hata et
al~(1995a) have argued that the discrepancy is real and is evidence
for new physics, e.g., an unstable tau neutrino of mass $10\MeV$
or so or neutrino chemical potentials.

Eventually the deuterium abundance in high
redshift Ly-$\alpha$ clouds will be decisive;  e.g. a value  (D/H)$_P\sim
10^{-4}$ implies $\eta \sim 2\times 10^{-10}$ and would implicate the
chemical evolution of $^3$He, while (D/H)$_P\sim 3\times 10^{-5}$
implies $\eta \sim 6\times 10^{-10}$ and would
implicate the primeval $^4$He abundance.  Until a definitive determination
is forthcoming, continued scrutiny of $^3$He -- both theoretically and
observationally -- offers a means of addressing this important issue.
Because previous measurements of the abundance of $^3$He have
raised as many questions as they have answered -- variations in
the abundance measured in HII regions of more than a factor of
five (Bania, Rood \& Wilson 1987; Wilson \& Rood 1994) 
with some values {\it lower than that
in the pre-solar nebula} (Black 1972; Geiss \& Reeves 1972) --
 the measurement of the $^3$He abundance in the local
ISM by Gloeckler \& Geiss (1996) is an important development.
We will use it to derive a lower bound on the baryon
density which is more empirically rooted and less sensitive to
the questionable aspects of $^3$He evolution.

\section{The Evolution of D + $^3$He}

According to conventional stellar modeling, low-mass stars ($M\la 2
M_\odot$) are net producers of
$^3$He and high-mass stars preserve at least 20\% or so of their $^3$He.
Integrating over a Salpeter mass function, Dearborn et al~(1986)  found a
mean $^3$He survival fraction of 0.8.  The arguments of Yang et al~(1984) and
others since (see e.g., Steigman and Tosi 1992, 1995) 
have been predicated upon this ``conventional wisdom.''

As mentioned above, there are reasons to remain skeptical.  Most
importantly, there is precious little observational evidence to support
this picture, with some recent observations apparently contradicting
it, cf. Scully et al~(1995).   A number of authors (e.g. Gough \& Weiss 1976;
Schmitt, Rosner, \& Bohn 1984;
Zahn 1992; Hogan 1995; Wasserburg, Boothroyd, \& Sackmann 1995;
Charbonnel 1994, 1995; Haxton, private communication) have discussed mixing
mechanisms by which $^3$He would be brought deep enough to be burned
and become depleted (to which we will refer collectively as
a ``solar spoon'').  Wasserburg, Boothroyd, \& Sackmann (1995) have
emphasized how such a mixing mechanism might explain carbon and oxygen
isotopic anomalies seen in certain AGB stars and in some meteoritic
grains (also see, Charbonel 1994, 1995; Weiss, Wagenhuber, \& Denissenkov 1996; 
Boothroyd \& Malaney 1996) and Haxton has suggested
that such a mechanism could lessen or even alleviate the
solar neutrino problem.

Finally, Copi et al~(1995c) have emphasized
how the heterogeneity of Galactic abundances
complicates attempts to infer primeval D and $^3$He abundances.
Heterogeneity arises because the Galaxy is not necessarily well mixed
and material in different regions
has experienced different histories of stellar-processing.  Starting with
the same primordial abundances, present local abundances can
vary by a factor of as much as two (see Fig.~1).
While the most recent
HST observations (Linsky et al~1993, 1995) now show at most
a 10\% variation in D/H within the local ISM,
earlier Copernicus and IUE observations showed a larger variation
in the local ISM (for a discussion of this point see
Ferlet and Lemoine 1996).  And of course, the local ISM could be relatively
homogeneous with the Galaxy inhomogeneous on larger scales.

The observational situation has its share of vagaries. The deuterium
abundance has only been measured in nearby regions of the Galaxy, along
several lines of sight in the local ISM and in the pre-solar nebula.  For
the pre-solar nebula, a deuterium abundance, (D/H)$_\odot = (2.7\pm0.5\,{\rm
sys}\pm 1\,{\rm stat}\,)\times 10^{-5}$, is inferred from the difference
of two measurements, the $^3$He abundance in the solar wind, which
reflects the sum of the pre-solar D + $^3$He (Geiss \& Reeves 1972), and the
$^3$He abundance measured in gas rich meteorites (Black 1972),
which reflects the pre-solar $^3$He abundance.
The higher pre-solar deuterium
abundance is consistent with its expected decline with time
due to stellar processing.
(Measurements of the deuterium abundance using deuterated molecules, both
in the solar system and throughout the Galaxy, shed little light as the
effects of chemical fractionation are expected to be very significant and
are difficult to disentangle.)

As mentioned above, the pre-solar abundance of $^3$He has been measured in
primitive meteorites, ($^3$He/H)$_\odot = (1.5\pm 0.2\pm 0.3)\times
10^{-5}$.  The present $^3$He abundance has also been measured within the
Galaxy, in a number of HII regions and in a planetary nebula by means of
the $^3$He$^+$ hyperfine line (Rood, Bania, \& Wilson 1992, 1995) and in a HB
star by Hartoog (1979).  The abundances in HII regions
range from ($^3$He/H)$_{\rm HII} = 10^{-5}$ to $6\times 10^{-5}$, suggesting
a wide variation in the present abundance.  On the
face of it, the planetary nebula measurement, ($^3$He/H)$_{\rm PN} \sim
10^{-3}$ and the HB star measurement are consistent with
the notion that low-mass stars produce significant amounts of
$^3$He.  However, only a few objects have been studied and these objects
represent a biased rather than representative sample -- optimized to
detect $^3$He (Rood, private communication).

Heterogeneity aside, the existing Galactic $^3$He measurements do not provide
a representative sample of material.  The
HII regions probably preferentially sample material that has been processed
through high-mass stars which destroy $^3$He (Olive et al~1995),
while the planetary nebulae and HB star represent objects with
sufficiently large $^3$He abundance to detect.

Finally, Gloeckler \& Geiss~(1996) have used the Solar Wind
Ion Composition Spectrometer (SWICS) on the Ulysses spacecraft
to determine the abundance of $^3$He in the local ISM, yielding a value 
($^3$He/H)$_{\rm ISM}
= (2.1^{+0.9}_{-0.8})\times 10^{-5}$, where the error
is the sum of statistical + systematic.  (They measured the
abundance of slowly moving, singularly ionized Helium atoms -- so-called
pick-up ions -- which entered the solar system as neutral atoms, were
photoionized and swept back out by the solar wind.)
This measurement is important because
the deuterium abundance in the ISM is also known.
Together they imply [(D + $^3$He)/H]$_{\rm ISM} = (3.7\pm 0.9)\times 10^{-5}$,
which is essentially identical to the pre-solar value,
[(D + $^3$He)/H]$_\odot \simeq (4.2 \pm 0.7 \pm 1)\times 10^{-5}$.
The constancy of the D + $^3$He
abundance over the past 4.5 Gyr is striking and provides
general confirmation of the cosmological utility of D + $^3$He
as proposed by Yang et al (1984), though one must be mindful of
the details of its implementation.

\section{Discussion}

Gloeckler \& Geiss' measurement is noteworthy because it
allows the evolution of D + $^3$He to be
addressed empirically for the first time. The message is simple:  over
the past 4.5 Gyr it has not changed dramatically.  This means that those
stars that have contributed significantly to the local ISM over this
period are not significant producers or destroyers of
$^3$He.  This is not a trivial fact, as the increase in $^3$He and the
decline in D over this time (almost a factor of two) indicate substantial
stellar processing.

According to chemical-evolution models, low-mass stars
($\sim 1M_\odot - 1.5 M_\odot$) have made the
dominant contribution to the ISM over the past 4.5 Gyr (Truran \&
Cameron 1971; Rood, Steigman \& Tinsley 1976; see also the recent
discussion by Scully et al~1995). Such a constancy of D + $^3$He
implies that low-mass stars cannot be significant producers of
$^3$He, which is at variance with the predictions of standard stellar
models.  Likewise,
there is no evidence to support significant destruction of $^3$He by
low-mass stars as predicted with an efficient solar spoon at work
(see e.g., Hogan 1995).  However, Dearborn (private conversation) has shown
that the slow mixing models that fit the oxygen and carbon
isotopic anomalies do not completely deplete $^3$He; they
reduce the amount of $^3$He that would have been returned
to the ISM by at most 80\%.  If this is indeed the case,
a solar spoon could be consistent with the ISM value of D + $^3$He.

Gloeckler \& Geiss' result does little to directly constrain the
earlier evolution of D + $^3$He.  The stellar mass
function at earlier times is expected to favor more massive stars,
which deplete $^3$He.  Since the Gloeckler \& Geiss result
constrains low-mass star destruction of $^3$He,
massive stars are the only possible way to greatly deplete $^3$He.
Massive stars produce heavy elements, and thus there is a limit
to the amount of material that could have been processed through massive stars.

In particular, the ejecta of Type II supernovae are about
10\% oxygen by mass, which implies
that only a small fraction of the material in the local ISM -- roughly
10\% -- could have been processed through massive stars.  Taken
together with the apparent constancy of D + $^3$He over the last 4.5
Gyr, this suggests that the primordial value of D +
$^3$He cannot differ greatly (about a factor of two for simple closed galaxy
models) from the present value.
This leads to the lower bound, $[({\rm D + ^3He})/{\rm H}]_P
\la 10^{-4}$ which is almost identical to that used by Yang et
al~(1984),  but now with firmer empirical roots.

An important assumption underlies the above argument,
that all the metals ejected by massive stars
make their way back into the ISM.  It is possible that metals
produced in the early supernova-active phase of the proto-Galaxy
(or the proto-galactessimals
that merged to form the Galaxy) were ejected into the surrounding IGM.
There is some evidence for this; observations of
the x-ray emitting gas in rich clusters show
metallicities that are slightly less than solar,
distributed in a gas mass that is roughly ten times that in
galaxies.  If the Galaxy ejected a comparable amount
of metals ten times more material could have been processed
through massive stars, depleting  $^3$He dramatically.
(Note, material depleted in $^3$He is still returned to
the ISM in a pre-supernova stellar wind.)  However, as Copi
et al~(1995c) showed,
even a relaxed metallicity constraint does not allow the primordial value
of $({\rm D + ^3He})/{\rm H}$ to exceed about $2 \times 10^{-4}$.

Finally, let us use the information gleaned from this first
measurement of the $^3$He abundance in the ISM to make more
quantitative statements about the value of the
baryon density and the consistency of standard big-bang
nucleosynthesis.  The stochastic history approach of
Copi et al (1995c) allows one to use the pre-solar values of $^3$He and D
+ $^3$He to infer both their primordial values and $\eta$, while allowing
for the heterogeneity of Galactic abundances.  The physical input
needed is the mean properties of stellar processing.  Based upon
Gloeckler \& Geiss' measurement, we consider two possibilities for
the evolution of $^3$He in low-mass stars, (1)  low-mass stars preserve their
$^3$He, but do not produce $^3$He; and (2) low-mass stars destroy 80\%
of the $^3$He they would have returned to the ISM, and two possibilities
for metal ejection by massive stars, (a) massive stars return
most of the metals they make to the ISM and (b)
massive stars only return 10\% of the metals they make
to the ISM (the rest ejected into the IGM).
For these four possibilities, 1a, 1b, 2a, and 2b, we have
constructed Monte-Carlo likelihood functions for the
baryon-to-photon ratio $\eta$, which are shown in Figure 2.
The 95\% credible intervals are: $\eta_{1a} = (5 - 7) \times
10^{-10}$; $\eta_{2a} = (3 -6) \times 10^{-10}$; $\eta_{1b}
= (2 - 5) \times 10^{-10}$; and $\eta_{2b} = (2 - 5) \times 10^{-10}$.
For reference, the very naive assumption that D + $^3$He
has remained unchanged since primordial nucleosynthesis
implies $\eta \sim 5 \times 10^{-10}$ and a primeval D
abundance $({\rm D/H})_P \sim 4\times 10^{-5}$.

Regarding the consistency of big-bang nucleosynthesis,
Model 1a continues to implicate
$^4$He as the culprit (or the standard model of big-bang nucleosynthesis
itself).  Models 1b, 2a, and 2b lessen the tension between
$^4$He and $^3$He and D, with Models 1b and 2b essentially eliminating
the tension all together.  The full range for the baryon density based
upon these models, $\eta \simeq (2-7) \times 10^{-10}$, is essentially
the same as that found previously by Copi et al (1995a).
We note that the models that lessen the tension,
lead to a stronger upper limit to $\eta$ and strengthen the case for
nonbaryonic dark matter.  For example, for Models 1b and 2b, the joint 95\%
credible region for all the light-elements
corresponds to $\Omega_B = (0.007 - 0.018)h^{-2}$.

In sum, the measurement of the interstellar $^3$He abundance
by Gloeckler \& Geiss (1996) allows the chemical evolution of D + $^3$He
to be addressed empirically for the first time, and in turn, tests
primordial nucleosynthesis and its prediction for the baryon density.
Their measurement indicates little evolution of D +
$^3$He over the past 4.5 Gyr, generally confirming the
the argument of Yang et al (1984), suggesting that low-mass stars are not
significant producers or destroyers  of $^3$He, and calling into question
standard stellar models as well as efficient solar spoons.
Little can be learned directly from their result about the earlier
evolution of D + $^3$He, which is likely to be dominated
by high-mass stars.  However, the fact that high-mass stars also
produce metals limits the amount of $^3$He depletion,
even if 90\% of the metals they produce are ejected from the Galaxy.
We have used this fact together with Gloeckler \& Geiss' result to
establish a more empirically based lower bound to the baryon-to-photon ratio,
$\eta \ga 2\times 10^{-10}$.  While only slightly less stringent
than the bounds of Yang et al~(1984) and Copi et al (1995a),
it suggests the apparent tension between the big-bang abundance of $^4$He
and those of D and $^3$He involves the chemical evolution of $^3$He.

\paragraph{Acknowledgments.} We acknowledge valuable conversations
with Robert Rood and John Simpson.  This work was supported in part by the DOE
(at Chicago and Fermilab) and the NASA (at Fermilab through grant NAG
5-2788 and at Chicago through NAG 5-2770 and a GSRP fellowship) and by NSF at
Chicago through grant AST 92-17969. 

\def\beginapjbib{\begingroup \section*{References}
         \parskip=.5ex plus 1.0pt 
	 \def\bibitem{\par \noindent \hangindent\parindent
		\hangafter=1}}
\def\endapjbib{\par \endgroup}

\beginapjbib

\bibitem Bania, T.M., Rood, R.T., \& Wilson, T.L. 1987, ApJ, 323, 30

\bibitem Black, D.C. 1972, Geochim. Cosmochim. Acta, 36, 347

\bibitem Boothroyd, A.I., \& Malaney, R.A. 1996, ApJ, in press

\bibitem Carswell, R.F., Rauch, M., Weymann, R.J., Cooke, A.J., \& Webb, J.K. 1994, MNRAS, 268, L1

\bibitem Carswell, R. F. et al~1995, MNRAS, in press

\bibitem Charbonnel, C. 1994, A\&A, 282, 811

\bibitem Charbonnel, C. 1995, ApJ, 453, L41

\bibitem Copi, C., Schramm, D.N., \& Turner, M.S. 1995a, Science, 267, 192

\bibitem Copi, C., Schramm, D.N., \& Turner, M.S. 1995b, PRL, 75, 3981.

\bibitem Copi, C., Schramm, D.N., \& Turner, M.S. 1995c, ApJ 455, L95.

\bibitem Dearborn, D.S.P., Schramm, D.N., \& Steigman, G. 1986, ApJ, 302, 35

%%\bibitem Dearborn, D.S.P., Steigman, G., \& Tosi, M. 1996, ApJ, in  press

\bibitem Epstein, R., Lattimer, J., \& Schramm, D.N. 1976, Nature, 263, 198 

\bibitem Ferlet, R. \& Lemoine, M. 1996, in {\em Cosmic Abundances;
Proceedings of the 6th Annual October Astrophysics Conference in Maryland}, in
press 

\bibitem Geiss, J., \& Reeves, H. 1972, A\&A, 18, 126

\bibitem Gloeckler, G. \& Geiss, J.~1996, Nature, submitted

\bibitem Gough, D.O. \& Weiss, N.O. 1976, MNRAS, 176, 589

\bibitem Hartoog, M. 1979, ApJ, 231, 161

\bibitem Hata, N., Scherrer, R.J., Steigman, G., Thomas, D., Walker, T.P., Bludman, S., \& Langacker, P. 1995, Phys Rev Lett, 75, 3977 

%%\bibitem Haxton, W. 1996, ??

\bibitem Hogan, C.J. 1995, ApJ, 441, L17

\bibitem Iben, I., Jr., \& Truran, J.W. 1978, ApJ, 220, 980

\bibitem Linsky, J.L., Brown, A., Gayley, K., Diplas, A.,
Savage, B.D., Ayres, T.R., Landsman, W., Shore, S.W., \& Heap, S. 1993,
ApJ, 402, 694

\bibitem Linsky, J. L., et al~1995, ApJ, in press

\bibitem Olive, K.A., \& Steigman, G. 1995, ApJS, 97, 49 

\bibitem Olive, K.A., Rood, R.I., Schramm, D.N., Truran, J.W., \& Vangioni-Flam, E. 1995, ApJ, 444, 680

%%\bibitem Pagel, B.~199?, ??

\bibitem Reeves, H., et al~1973, ApJ, 179, 909

\bibitem Rood, R.T., Bania, T.M. \& Wilson, T.L. 1992, Nature, 355, 618

\bibitem Rood, R.T., Bania, T.M. \& Wilson, T.L. 1995, in {\it Light Element Abundances; Proceedings of the ESO/EIPC Workshop}, ed. P. Crane (Berlin:Springer), p. 201

\bibitem Rood, R.T., Steigman, G., \& Tinsley, B.M. 1976, ApJ, 207, L57

\bibitem Rugers, M. \& Hogan, C. J.~1996, ApJ, 459, L1

\bibitem Sasselov, D. \& Goldwirth, D. S.~1995, ApJ, 444, L5

\bibitem Schmitt, J.H.M.M., Rosner, R., \& Bohn, H.U. 1984, ApJ, 282, 316

\bibitem Scully, S.T., Cass\'e, M., Olive, K.A., Schramm, D.N.,
Truran, J.W., \& Vangioni-Flam, E. 1996, ApJ, in press

\bibitem Skillman, E.D., Terlevich, R.J., Kennicutt, R.C., Garnett, D.R., \&
Terlevich, E. 1994, ApJ, 431, 172

\bibitem Songaila, A., Cowie, L.L., Hogan, C.J., \& Rugers, M. 1994, Nature, 368, 599

\bibitem Steigman, G., \& Tosi, M. 1992, ApJ, 401, 150

\bibitem Steigman, G., \& Tosi, M. 1995, ApJ, 453, 173

\bibitem Truran, J.W., \& Cameron, A.G.W. 1971, ApSpSci, 14, 179

\bibitem Tytler, D. \& Fann, X. M.~1994, BAAS, 26, 4, 1424

\bibitem Walker, T.P., Steigman, G., Schramm, D.N., Olive, K.A., \& Kang, H. 1991, ApJ, 376, 51

\bibitem Wampler, E. J., et al~1996, A\&A, in press

\bibitem Wasserburg, G.J., Boothroyd, A.I., \& Sackmann, I.-Juliana 1995, ApJ, 447, L37

\bibitem Weiss, A., Wagenhuber, J., \& Denissenkov, P.A. 1996, A\&A, in press

\bibitem Wilson, T.L. \& Rood, R.T. 1994, ARAA, 32, 191

\bibitem Yang, J., Turner, M.S., Steigman, G., Schramm, D.N., \& Olive, K.A. 1984, ApJ, 281, 493

\bibitem York, D., et al~1984, ApJ, 276, 92.

%\bibitem York, D., et al~1995, ??

\bibitem Zahn, J.P. 1992, A\&A, 265, 115

\endapjbib

%\newpage

%\section*{Figure Captions}

%\bigskip
%\noindent{\bf Figure 1:}  
\begin{figure} \epsfysize=6.5in
\center \leavevmode 
\rotate[r]{\epsfbox{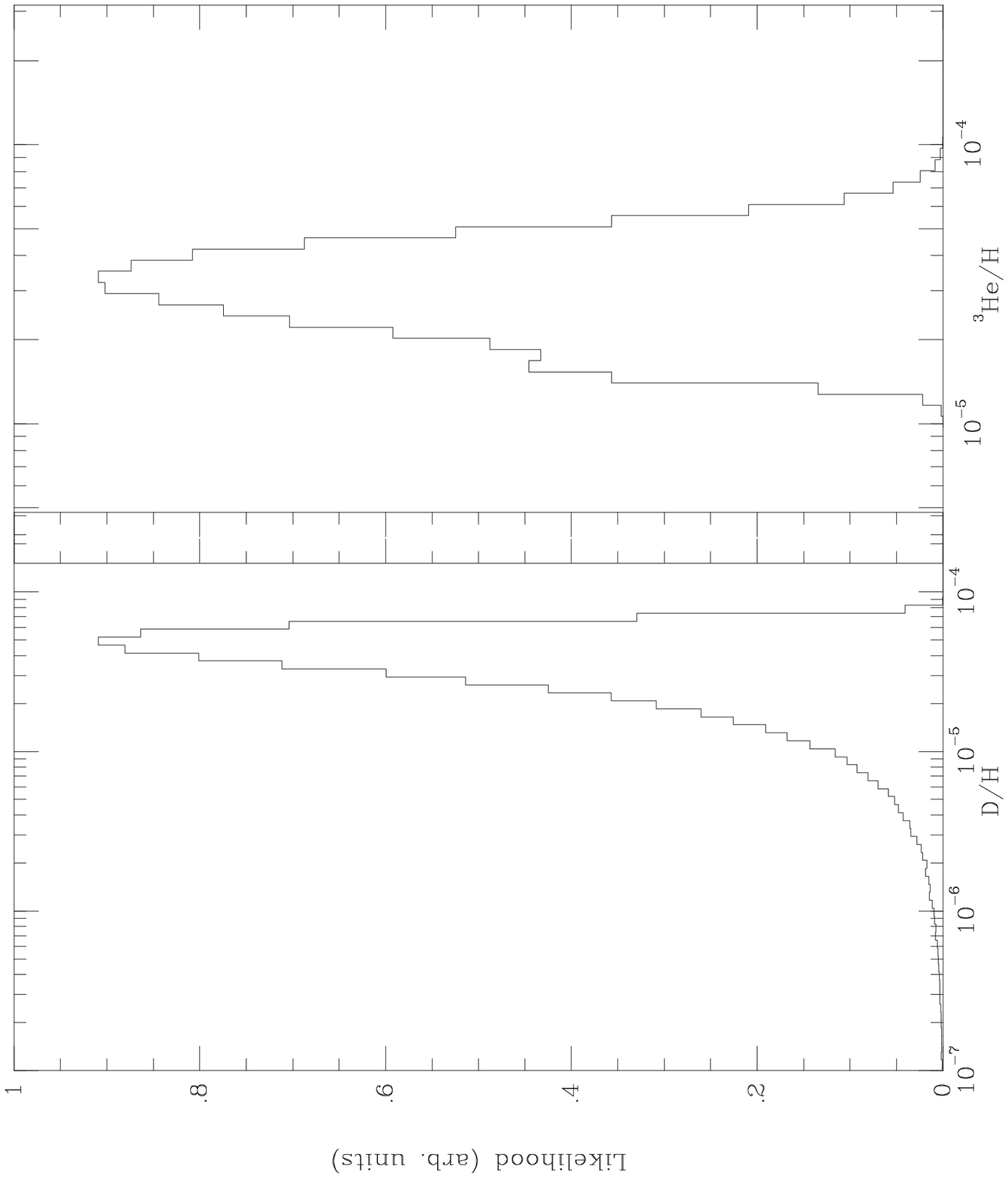}}
\caption{
The variation in present D and $^3$He abundances
expected today due to differing histories.  Here we have assumed
$\eta=3.2\times 10^{-10}$ to fix the primordial values.
}
\end{figure}

%\medskip
%\noindent{\bf Figure 2:}
\begin{figure} \epsfysize=6.5in
\center \leavevmode
\rotate[r]{\epsfbox{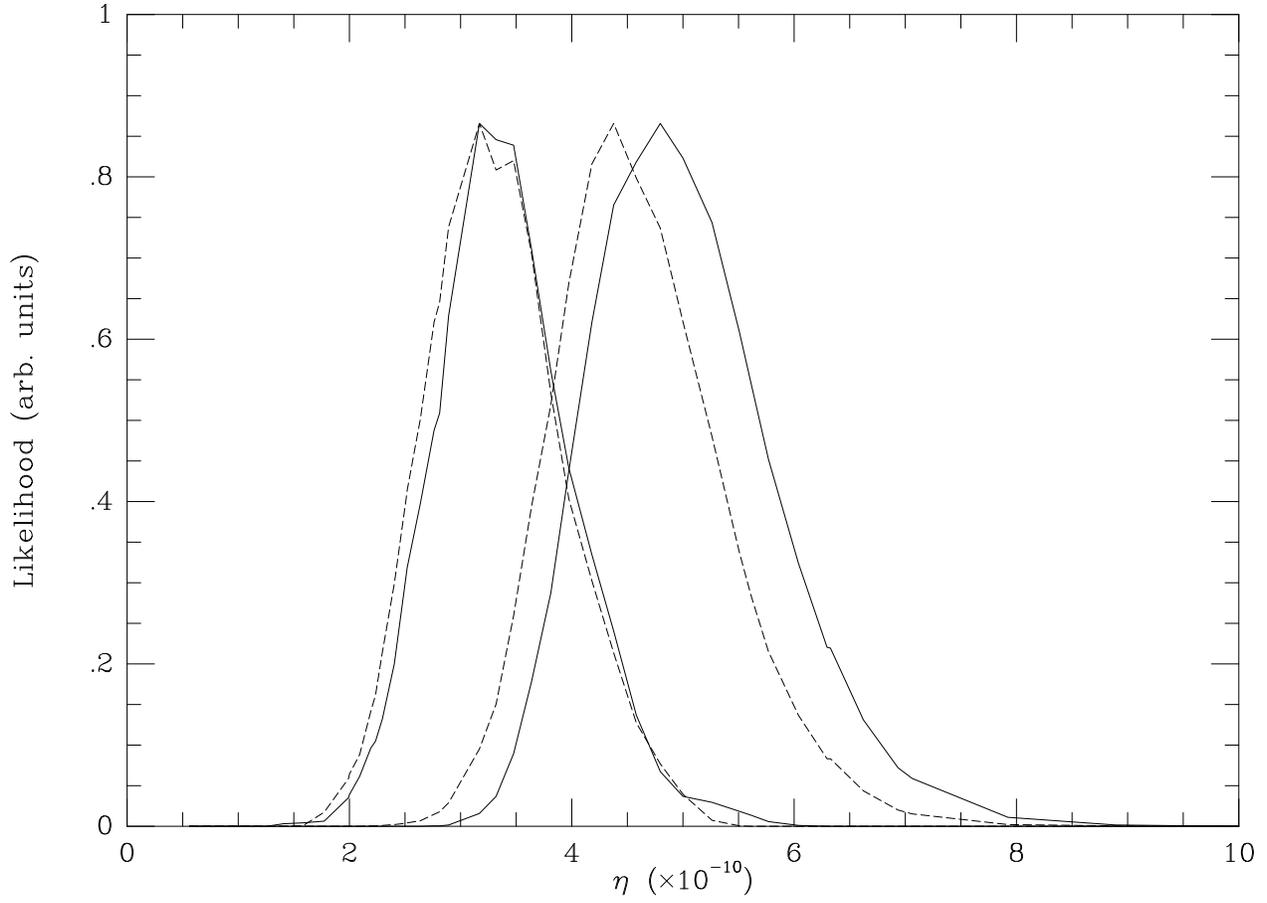}}
\caption{
  Monte-Carlo likelihood functions
for the baryon-to-photon ratio based upon D and $^3$He for
Models 1a (right solid curve), 2a (right broken curve), 1b (left solid curve),
and 2b (left broken curve).
}
\end{figure}

\end{document}